\documentclass[useAMS]{mn2e}
\usepackage{psfig}
\usepackage{times}

\def\spose#1{\hbox to 0pt{#1\hss}}
\newcommand\lsim{\mathrel{\spose{\lower 3pt\hbox{$\mathchar"218$}}
     \raise 2.0pt\hbox{$\mathchar"13C$}}}
\newcommand\gsim{\mathrel{\spose{\lower 3pt\hbox{$\mathchar"218$}}
     \raise 2.0pt\hbox{$\mathchar"13E$}}}

\title[Absorption of $\gamma$-rays in 3C279 and EBL constraints]
{Intrinsic absorption in 3C 279 at GeV-TeV energies and consequences
for estimates of the EBL}

\author[Tavecchio \& Mazin] {Fabrizio Tavecchio$^1$\thanks{E--mail:
fabrizio.tavecchio@brera.inaf.it} and Daniel Mazin$^2$\\ 
$^1$ INAF -- Osservatorio Astronomico di Brera, via E. Bianchi 46,
I--23807, Merate, Italy\\
$^2$ IFAE, Edifici Cn., Campus UAB, E-08193 Bellaterra, Spain.}

\begin{document}

% \date{Accepted 1988 December 15. Received 1988 December 14; 
% in original form 1988 October 11}

\pagerange{\pageref{firstpage}--\pageref{lastpage}} \pubyear{2007}

\maketitle

\label{firstpage}

\begin{abstract} 
We revisit the limits of the level of the extragalactic background
light (EBL) recently reported by the MAGIC collaboration based on the
observed $\gamma$-ray spectrum of the quasar 3C279, considering the
impact of absorption of high-energy $\gamma$-ray photons inside the
broad line region (BLR) of the quasar. We use the photoionization code
CLOUDY to calculate the expected optical-UV radiation field inside the
BLR and the optical depth to $\gamma$-rays for a relatively extended
set of the parameters. We found that the absorption of $\gamma$-ray
photons, though important for the estimate of the true radiative
output of the source, does not produce an important hardening of the
spectrum of 3C279 in the energy band accessible by MAGIC, supporting
the method used to infer the upper limits to the level of the EBL.
\end{abstract}

\section{Introduction}

The detection of 3C279 ($z$=0.536) in the very high energy (VHE,
$E>50$\,GeV) band by the MAGIC telescope (Albert et al. 2008) extends
to the quasars the group of the known extragalactic VHE sources,
before limited to BL Lac objects (excluding the nearby radiogalaxy
M87)\footnote{see {\tt
http://www.mppmu.mpg.de/$\sim$rwagner/sources}}.

For some aspects, the detection of 3C279 comes as a surprise. General
theoretical arguments support the view that quasars cannot be
important VHE emitters, in particular because of the expected
absorption, through the pair production process $\gamma + \gamma
\rightarrow e^+ + e^-$, inside the source itself (among the most
recent calculations, e.g., Donea \& Protheroe 2003, Liu \& Bai 2006,
Reimer 2007). Moreover, quasars are generally located at a relatively
high redshift, implying a huge absorption by the extragalactic
background light (EBL, e.g. Kneiske et al. 2004, Primack et al. 2005,
Stecker et al. 2006, Franceschini et al. 2008).  Last, but not least,
widely adopted standard leptonic models for production of $\gamma
$-rays in 3C279 (Hartman et al. 2001, Ballo et al. 2002; see also
Tavecchio \& Ghisellini 2008, hereafter TG08) do not predict an
important emission above few tens of GeV, because of the rapid
decrease of the scattering cross section (for hadronic scenarios see,
e.g. Mannheim 1993). The observation of VHE photons from 3C279 by
MAGIC (Albert et al. 2008) demonstrates that also quasars can, to some
extent, produce high-energy $\gamma $-rays and suggests that the
opacity (both intrinsic and cosmic) is less strong than what
previously assumed.

The detection of a source of VHE photons at a relatively high redshift
offers a unique tool to probe the still poorly known EBL. Albert et
al. (2008) used general arguments based on a limiting spectral slope
for the emitted spectrum to infer the amount of extragalactic
absorption\footnote{See Stecker \& Scully (2008) for an alternative
interpretation.}. However, as recently pointed out by Aharonian et
al. (2008) (see also Bednarek 1997), intrinsic absorption of
$\gamma$-ray photons inside the source could result in rather hard
observed spectra. Such spectra affected by absorption could lead to
severely underestimate the level of the EBL when the arguments based
on the hardness of the spectrum are used. For this reason, the use of
the measured spectrum to constrain the EBL in Albert et al. (2008)
triggered some discussion on the role and strength of the internal
absorption in this source (Liu et al. 2008, Sitarek \& Bednarek 2008).

We emphasize that the absorbed spectrum will be harder than the
intrinsic one only in the case of an optical depth $\tau(E)$ {\it
decreasing} with energy, $E$. Thus, as long as the optical depth
increases (or, at least, only slightly decreases) with energy, the
resulting spectrum will be softer (or slightly harder) than the
intrinsic one and the standard spectral methods to constrain the EBL
could be safely used. It is thus clear that the shape of $\tau(E)$,
strictly related to the spectrum of the target photons, is the key
element to assess the effects of absorption on these methods.

Previous attempts to calculate intrinsic absorption in the BLR (Liu \&
Bai 2006, Reimer 2007, Liu et al. 2008) assumed rather idealized
templates for the BLR spectrum, considering the most prominent
emission lines but neglecting the important contribution of the
optical-UV continuum. In a recent paper, Sitarek \& Bednarek (2008)
include the absorption in a self-consistent model for the high-energy
emission of 3C279. They consider in detail the geometry of the
radiation fields inside the BLR, including also the contribution of
direct and scattered radiation of the accretion disc, but, again, they
model the BLR radiation with the simplified template of Liu \& Bai
(2006).

In this work (Sect. 2) we explore the effects of absorption using more
realistic spectra of the BLR calculated with the photoionization code
{\tt CLOUDY} (Ferland et al. 1998), previously used to study in detail
the inverse Compton emission from powerful blazars (TG08). In
particular, we show that the IR-optical-UV continuum plays an important
role in determining the absorption, resulting in an almost constant
optical depth in the broad energy range 30 GeV--30 TeV. In Sect. 3 we
use the spectra corrected for absorption to revisit the constraints on
the EBL based on the spectrum of 3C279 derived in Albert et
al. (2008). In Sect. 4 we discuss the results.

\section{Intrinsic absorption}

\subsection{The model}

\begin{figure}
%\vspace*{-1.9 truecm}
{\hspace*{-.4 truecm}
\psfig{file=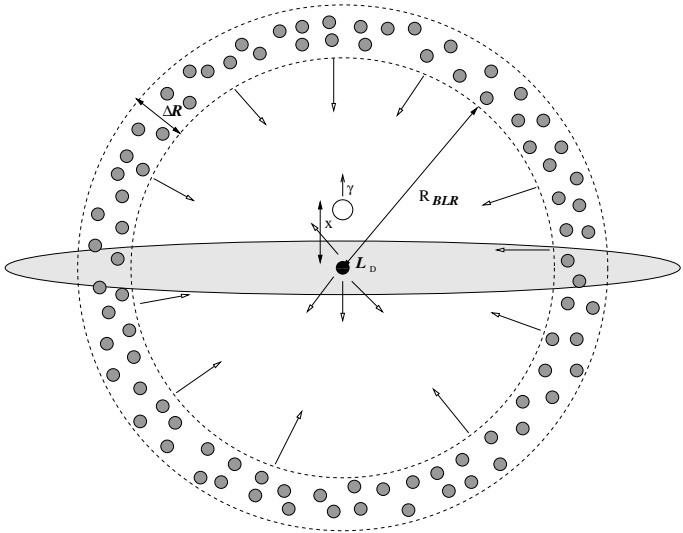,width=9.cm}}
\caption{Sketch of the geometry assumed in the model (not to
scale). The (uniform) BLR is assumed to be a spherical shell with
thickness $\Delta R$ and inner radius $R_{\rm BLR}$, illuminated by
the central continuum with luminosity $L_{\rm D}$. The source of
$\gamma$-ray photons is at distance $x$ from the central black
hole. See text for details.}
\label{cartoon}
\end{figure}

We calculate the diffuse radiation field inside the BLR of 3C279 for
different values of the BLR radius, temperature of the accretion disk,
slope of the illuminating UV radiation. We refer to TG08 for a full
description of the model. We assume the geometry shown in
Fig.\ref{cartoon}. The accretion flow illuminates the BLR clouds
(characterized by the total hydrogen density $n$ and the hydrogen
column density $N_H$) isotropically filling the BLR, assumed to be a
thin spherical shell with inner radius $R_{\rm BLR}$. In the
calculation we assume that the clouds cover a fraction
$C=\Omega/4\pi=0.1$ of the solid angle viewed from the central
illuminating source. The emission from the illuminated face of the
clouds is calculated with version 05.07 of {\tt CLOUDY}, described by
Ferland et al. (1998)\footnote{See also {\tt http://www.nublado.org}}.
For simplicity, we discuss only the case of solar abundance and in all
calculations we fix $n=10^{10}$ cm$^{-3}$ and $N_H=10^{23}$
cm$^{-2}$. Results do not substantially change for different densities
and column densities (see TG08). We adopt the spectrum of the
illuminating continuum modeled as a combination of a UV bump (with
slope $\alpha _{UV}$) with a flat X-ray power-law, $L_D(\nu) \propto
\nu ^{-1}$, commonly assumed in these calculations ({\tt AGN} model in
{\tt CLOUDY}, e.g. Korista \& Goad 2001 and references therein). When
not explicitly noted, the disk is assumed to have the ``standard''
temperature $T_D=1.5\times 10^5$ K.

We assume that the disk in 3C279 emits a total luminosity $L_D=2\times
10^{45}$ erg/s (Pian et al. 1999). More uncertain is the radius of the
BLR. The empirical relations connecting the luminosity of the disk and
$R_{\rm BLR}$ (Bentz et al. 2006, Kaspi et al. 2007) provide $R_{\rm
BLR}\simeq 1-3 \times 10^{17}$ cm. Given these uncertainties, below we
show the results for four different values of $R_{\rm BLR}$, $1,
1.6, 3.2 \;{\rm and}\; 6.3\times 10^{17}$ cm.

The optical depth for the photon-photon absorption, $\tau (E)$, is
calculated as (e.g., Liu \& Bai 2006):
\begin{equation}
\tau(E)=\int_{x}^{R_{BLR}} \int \int
n(\nu,\Omega,l) \sigma _{\gamma\gamma}(E,\nu,\Omega) (1-\mu )
d\Omega  d\nu  dl
\label{tau}
\end{equation}
\noindent
where $E$ is the energy of the $\gamma $-rays, $l$ is the distance of
the photons from the BH, $n(\nu,\Omega,l)$ is the number density of
the radiation for solid angle at each location of the photon path,
$\sigma _{\gamma\gamma}(E,\nu,\Omega)$ is the cross section and
$d\Omega =2\pi d\mu$.  Note that, given the characteristics of the
cross section, for a fixed geometry $\tau (E) \propto \nu n(\nu) $
with $\nu \propto 1/E$, that is the optical depth reflects the
spectrum of the soft photon field (smeared by the cross section).

To calculate $\tau(E)$ from Eq.(\ref{tau}) one has to specify the
location of the source, $x$.  All calculations discussed below have
been performed with $x=2\times 10^{16}$ cm. Larger values of $x$ (a
source closer to the BLR) result in a lower level of the absorption,
but do not essentially affect the {\it shape} of $\tau(E)$. For
smaller values of $x$ one should also consider the absorption induced
by photons directly coming from the accretion disk (Ghisellini \&
Madau 1996, Sitarek \& Bednarek 2008). However, absorption from the
disk is characterized, for the considered range of energies, by an
optical depth monotonically increasing with energy (Sitarek \&
Bednarek 2008).  As already discussed, in this conditions the
resulting spectrum is {\it softer} than the emitted spectrum and thus
it does not affect the constraints on the EBL.

Note that, for simplicity, we neglect the effects related to the
radiative transfer {\it inside} the emission region (negligible as
long as the size of the source is much less than $R_{\rm
BLR}-x$). Another possible effect that we do not take into account is
that in the case of a moving source (as in the standard ``internal
shock'' scenario, Spada et al. 2001), photons emitted at different
times will also be characterized by different $x$ in Eq.\ref{tau} and
thus will suffer a different level of intrinsic absorption. The study
of these effects, in part already considered by Sitarek \& Bednarek
(2008), is important in view of the interpretation of the
time-resolved spectra soon available thanks to the {\it Fermi
Gamma-ray Space Telescope}.

\subsection{Results}

\begin{figure}
\vspace*{-1 truecm}
{\hspace*{-1.1 truecm}
\psfig{file=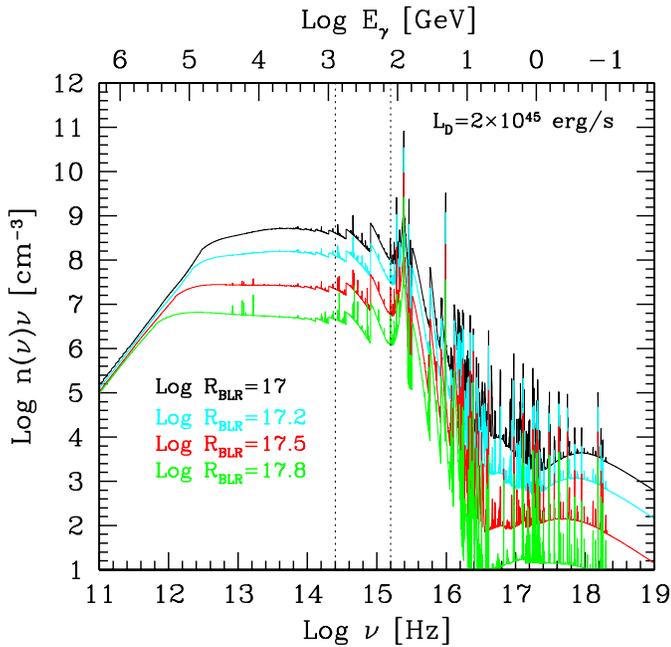,height=11cm,width=11cm}}
\vspace*{-1.2 truecm}
\caption{Number density of BLR photons for $L_D=2\times 10^{45}$
erg/s, $T_D=1.5\times 10^5$ K, $\alpha _{UV}=0.5$ and BLR radii of 1,
1.6, 3.2 and $6.3 \times 10^{17}$ cm (cases S in Fig.3). Note the
important component in the optical-UV band, mainly coming from from
free-free emission, Thomson and Rayleigh scattering. The upper
$x-$axis reports the energy of $\gamma$-rays mainly absorbed by soft
photons of the corresponding frequency. The two dotted vertical lines
show the frequency range (in the quasar rest frame) interesting for
absorption of $\gamma$-rays in the interval of energies covered by the
MAGIC spectrum reported by Albert et al. (2008).}
\label{lines}
\end{figure}

As an example, Fig.\ref{lines} reports some BLR spectra derived with
the model, assuming $T_D=1.5\times 10^5$ K, $\alpha _{UV}=0.5$,
$R_{\rm BLR}$, $1, 1.6,\ 3.2\;{\rm and}\; 6.3\times 10^{17}$ cm,
plotted as the number photon density, $n(\nu)\nu$. The two vertical
lines indicate the spectral range interesting for absorption of
$\gamma$-rays in the energy range covered by the MAGIC spectrum (upper
$x$-axis, in the quasar rest frame). Clearly, besides the emission
lines, the continuum (deriving from free-free emission, recombinations
and scattering, both Thomson and Reyleigh) provides an important
contribution especially at low frequencies, below the Ly${\alpha}$
line. In particular, in the UV band the dominant contribution
comes from Thomson and Rayleigh scattering (e.g. Korista \& Ferland
1998), whereas at IR-optical frequencies (in the range $\nu =
10^{12}-10^{14}$ Hz), the emission is generally dominated by optically
thin free-free emission plus a contribution from Thomson scattering. The
result is that the $n(\nu)\nu$ curves are almost flat below the UV
band, translating in an almost constant, or slightly increasing,
optical depth for a rather broad interval of energies.

\begin{figure}
\vspace*{-1. truecm}
{\hspace*{-1.75 truecm}
\psfig{file=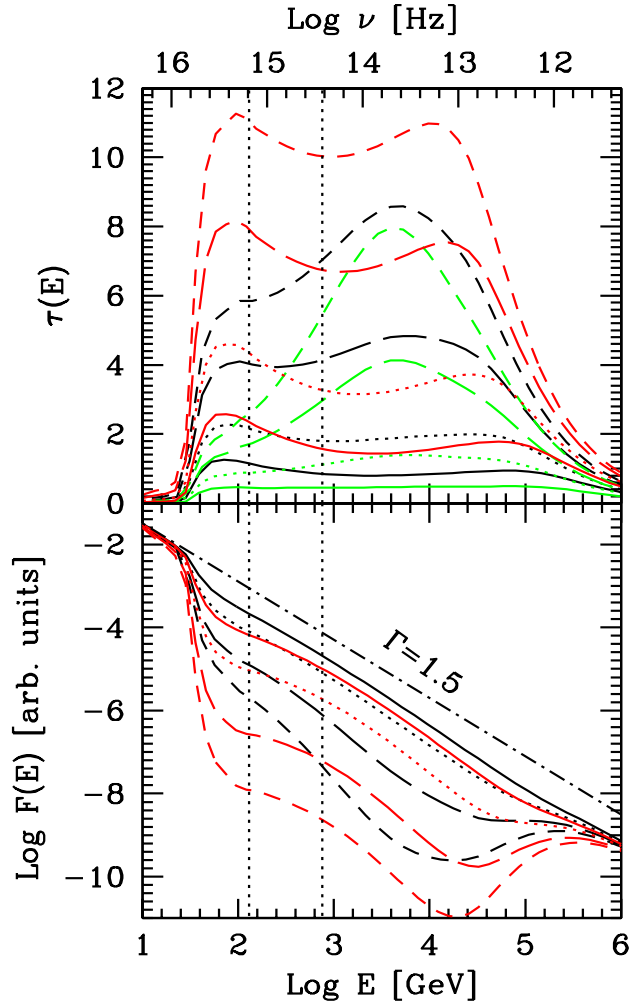,height=16cm,width=11.5cm}}
\vspace*{-1.2 truecm}
\caption{{\it Upper panel:} optical depth for absorption of
$\gamma$-rays for different parameters. Short dashed, long dashed,
dotted and solid lines: $R_{\rm BLR}=1, 1.6, 3.2\; {\rm and}\;6.3
\times 10^{17}$ cm, respectively. Black lines (case S): $\alpha
_{UV}=0.5$ and $T_D=1.5\times 10^5$ K. Red lines (case E): $\alpha
_{UV}=-1/3$. Green lines (case L): $T_D=5\times 10^4$ K. The two
vertical lines indicate the energy range of the observed spectrum of
3C279 (quasar rest frame). The upper $x$-axis shows the frequency of
the target photons mainly interacting with $\gamma$-rays of the energy
indicated in the lower $x$-axis. {\it Lower panel:} modification of
the intrinsic photon spectrum (assumed to be a power-law with slope
$\Gamma=1.5$, black dashed-dotted line) by absorption. For simplicity
we do not report the curves of the L case, always leading to a
softening of the spectrum in the MAGIC band. Lines styles and colours
as above.}
\label{abs}
\end{figure}

Fig.\ref{abs} illustrates some examples of the optical depth,
$\tau(E)$ (upper panel) and the corresponding absorbed spectra,
assuming an intrinsic photon spectrum $F(E)\propto E^{-1.5}$ (lower
panel). Different line styles and colours refer to different
parameters. Short dashed, long dashed, dotted and solid lines are for
$R_{\rm BLR}$, $1, 1.6, 3.2\; {\rm and}\; 6.3\times 10^{17}$ cm,
respectively. Black lines refer to the ``standard'' (S) scenario, with
$\alpha _{UV}=0.5$ (Elvis et al. 1994) and $T_D=1.5\times 10^5$ K. Red
lines (``extreme'' case, E) are calculated for a somewhat extreme
value of the UV slope, $\alpha _{UV}=-1/3$ (slope expected from a
standard thin disk), while green lines report the results for a ``low
temperature'' case (L), $T_D=5\times 10^4$ K (and $\alpha _{UV}=0.5$).

%--------------------------------------------------
\begin{figure*}
{\hspace*{-1.1 truecm}
\psfig{file=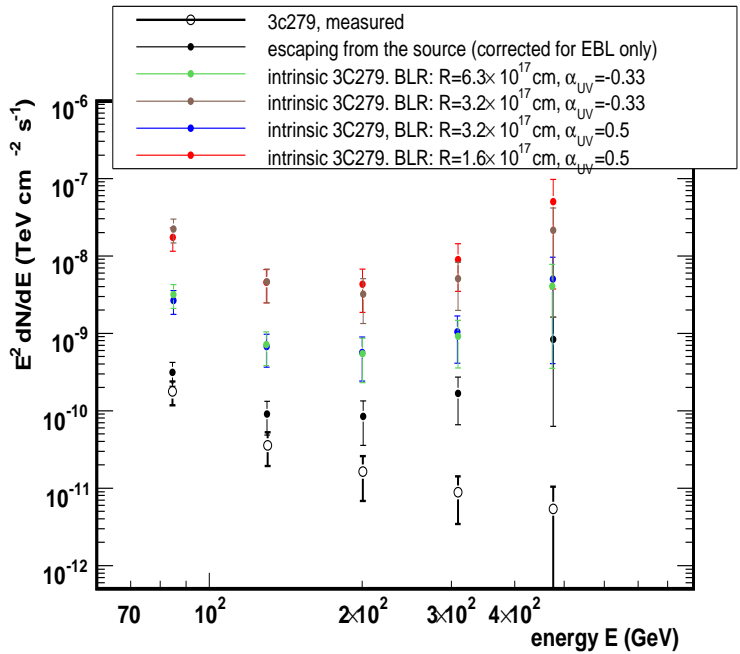,height=9.cm,width=10cm}}
\caption{Energy spectrum of 3C279. The measured spectrum (open
circles), EBL-corrected (filled black circles) and four different
scenarios for the internal absorption (in addition to the external
absorption) are shown.}
\label{3c279spec}
\end{figure*}

A common feature of these curves is the sudden increase of $\tau(E)$
starting around $E\simeq 20-30$ GeV, energy at the threshold for
photons of the Ly${\alpha}$ line. After reaching the maximum around
100 GeV, $\tau(E)$ displays a decreasing branch, just centered on the
energy range covered by the MAGIC spectrum (vertical lines). Without
the contribution of the continuum at frequencies below the
Ly${\alpha}$ line, the optical depth would fast decrease, determining
a very hard observed spectrum above 100 GeV (rest frame), as in the
model of Aharonian et al. (2008). However, the important contribution
of the continuum produces a bump extending above the TeV band,
hampering an important hardening of the out-coming $\gamma$-ray
spectrum (Fig.\ref{abs}, lower panel).

Some general characteristics of the optical depth in the
``line-dominated'' (below 100 GeV) and the ``continuum dominated''
(above 100 GeV) regions are clearly evident in these curves.  The
``line-dominated'' bump is strongly depressed in the L case. This is
due to the small fraction of photons above the Ly$\alpha$ energy, with
the subsequent depression of the luminosity of the Ly$\alpha$ line. In
this case the optical depth is always increasing with energy in the
interesting energy range. Another trend is visible by comparing the
curves corresponding to the S and the E case: in the latter the
`line-dominated'' component is systematically more prominent than in
the S case. The reason is that, for the same luminosity, the hard
illuminating continuum in the E case has a larger fraction of photons
above the Ly$\alpha$ energy than in the S case. A third trend visible in
Fig.\ref{abs}, especially for the L and the S curves, is the
increasing role of the ``continuum dominated'' region when $R_{\rm
BLR}$ decreases. The reason of this effect is the increasing number of
free electrons available for scattering and free-free emission due to
the increasing ionization at small radii. For the E case the
importance of this effects is partly reduced, because of the paucity
of the small optical-IR flux of the illuminator.

As we have already stressed, the hardening of the spectrum is realized
for decreasing values of the optical depth with energy. The key
parameter determining the slope of the optical depth (and thus the
slope of the modified $\gamma$-ray spectrum) at energies above $\sim
100$ GeV is the ratio between the optical depth in the
``line-dominated'' bump and in the ``continuum dominated''
region. This is a robust consequence of the realistic models of the
BLR radiation considered here. Therefore the hardest spectra will be
observed in the cases of the largest value of the ratio
line/continuum. As discussed above these conditions are realized for
large temperatures of the disk and hard UV slopes. Case E can then
be considered a conservative upper limit for the calculations.

Besides the spectral modifications, an important aspect to consider is
also the level of the absorption. In general, smaller radii imply
larger absorption, since $\tau \propto 1/R$. In the case of the
smaller radius considered here, $R_{BLR}=10^{17}$ cm, fluxes are
depressed by more than 2 orders of magnitude for the S and the E
cases, pushing the power requirements of the source to above $10^{50}$
erg/s. The absorption in the MAGIC band is instead limited in the case
of the L case. For the largest radius, $R_{BLR}=6.3\times 10^{17}$
cm, the requirements are still large for the E case ($\tau\sim 2$),
while for the other two cases the absorption is modest.

\section{Limits to the EBL}

In the following we discuss the effect of absorption on the
constraints for the EBL. We use the EBL model from Kneiske et
al. (2004), modified in Albert et al. (2008) to represent an upper
limit of the EBL level, which is in the same time a lower limit on the
transparency of the universe to VHE $\gamma$-rays. Using this
particular EBL model we examine different scenarios of the internal
absorption discussed above and revisit a possibility to emit
corresponding VHE spectra. From the discussion above it is
clear that the modification of the spectrum in the E case should be
considered a conservative upper limit to the real case. The L case is
not considered since it always leads to softer spectra. Moreover we
conservatively consider only the cases for which the ratio of the
absorbed and the emitted flux in the MAGIC band is larger than
$10^{-2}$ (corresponding to $\tau < 4.5$).

For the discussed scenarios of the internal absorption, we
reconstructed the intrinsic spectrum of 3C279. First, we corrected the
measured energy spectrum (Albert et al. 2008) for a given EBL model,
corresponding to the maximum allowed level derived in that paper. In
such a way we reconstruct the energy spectrum which escapes the
vicinity of 3C279 towards the observer. In the next step, we corrected
this spectrum for the intrinsic absorption to obtain the original
(produced) energy spectrum.  Results are shown in
Fig.\ref{3c279spec}. As discussed above, it can be seen that different
scenarios for the internal absorption mainly affect the flux level of
the emission but not the shape of the spectrum. Adopting the test
prescription from Mazin \& Raue (2007), we tested the resulting
intrinsic spectra for the criterion of the hardness ($\Gamma>1.5$) and
found that in all tested cases the intrinsic spectra can be excluded.
Consequently, identical or even harder EBL limits can be derived as
compared to the ones obtained in Albert et al. (2008).  Note that for
the tests not only the fit value of the slope from a simple power law
(PL) was examined but also slope values from a fit by a broken power
law (BPL). The latter one was tested in case the BPL fit gave a
significantly smaller $\chi^{2}$ value\footnote{criterion was defined
using the likelihood ratio test, for details see Mazin \& Raue (2007)}
than the PL.

\section{Discussion}

The detection of VHE $\gamma$-rays from 3C279 and EBL constraints
derived in Albert et al. (2008) triggered some discussion on the role
and strength of the internal absorption in this object (Liu et
al. 2008, Sitarek \& Bednarek 2008). In this paper, we have calculated
various models for the BLR diffuse radiation and applied them for the
case of 3C279.  We used the code {\tt CLOUDY} to calculate the models
in order to have a realistic energy spectrum of the photons inside of
the BLR region. The main difference between our model and those
of Liu et al. (2008) and Sitarek \& Bednarek (2008) is the assumed BLR
spectrum.  Liu et al. (2008) assume that the BLR spectrum consists
only of narrow lines. Sitarek \& Bednarek (2008) present a detailed
model for the high-energy emission of 3C279, including also the
possible contribution of direct and scattered disc radiation. However,
their spectrum of the BLR emission does not include other important
contributions to the continuum, such as the free-free emission.  As we
have shown, our model provides a strong continuum component, extending
on the optical-UV band (TG08).

Our results confirm that radiation from the BLR modifies the
primary emission of 3C279. However, we find that the internal
absorption inside the BLR does not produce an important hardening of
the spectrum in the energy band covered by the MAGIC observation. In
particular, we found that for the tested BLR models, despite a
possible overall softening of the 3C279 spectrum, at least part of the
spectrum was significantly above an implied maximum hardness of
$\Gamma=1.5$, confirming the EBL constraints derived in Albert et
al. (2008). Of course, the consideration of intrinsic absorption
implies that the TeV emission from 3C279 could be substantially more
powerful than published.

%We note that the 3C279 spectrum, corrected for the EBL and internal
%absorption, was submitted to fits more complicated than just a simple
%power law for the total energy range observed. 

Note also that conclusions from Sitarek \& Bednarek (2008) that an
EBL model of Stecker et al. (2006) does not imply an unrealistic
intrinsic spectra of 3C279 (when taking into account internal
absorption in the BLR) concern the ``baseline'' EBL model of Stecker
et al. (2006).  The EBL limits in this paper and in Albert et
al. (2008) instead concern the ``fast evolution'' model of Stecker et
al. (2006), implying a significantly higher EBL level in the redshift
range between $z=0$ and $z=1$.

We finally note that the discussion on the role of absorption
implicitly assumes that the highly variable high-energy $\gamma$-ray
emission detected from 3C279 is produced internally to the BLR
(probably through the comptonization of the BLR photons). However,
given the small size of the BLR in 3C279 as estimated from the
empirical relations connecting it to the disk luminosity, it is
conceivable that the emission is (at least in some occasions) produced
outside the BLR, thus avoiding the problems connected to
absorption. In this case, the mechanism responsible for the production
of the observed emission cannot be the external Compton: alternatives
include synchrotron self-Compton emission (Lindfors et al. 2006) or
comptonization of the IR radiation from the putative dusty torus
surrounding the central regions (e.g. Sikora et al. 2002, 2008;
Sokolov \& Marscher 2005).

\section*{Acknowledgements}
F.T. is grateful to Gabriele Ghisellini for enlightening discussions
and to Laura Maraschi for the critical and constructive reading of the
paper. We also thank Elina Lindfors for comments. We thank Gary
Ferland for maintaining his freely distributed code CLOUDY. D.M.'s
research was supported by a Marie Curie Intra European Fellowship
within the 7th European Community Framework Programme.

\end{document}